\documentclass[aps,pra,preprint,preprintnumbers,amsmath,amssymb]{revtex4}
\usepackage{graphicx,epsfig,epsf}
\usepackage{dcolumn}
\usepackage{bm}

\begin{document}
\newcommand{\ri}{{\rm i}}
\newcommand{\re}{{\rm e}}
\newcommand{\bx}{{\bf x}}
\newcommand{\bd}{{\bf d}}
\newcommand{\br}{{\bf r}}
\newcommand{\bk}{{\bf k}}
\newcommand{\bE}{{\bf E}}
\newcommand{\bR}{{\bf R}}
\newcommand{\bM}{{\bf M}}
\newcommand{\bn}{{\bf n}}
\newcommand{\bs}{{\bf s}}
\newcommand{\tbs}{\tilde{\bf s}}
\newcommand{\rSi}{{\rm Si}}
\newcommand{\beps}{\mbox{\boldmath{$\epsilon$}}}
\newcommand{\rg}{{\rm g}}
\newcommand{\tr}{{\rm tr}}
\newcommand{\xmax}{x_{\rm max}}
\newcommand{\ra}{{\rm a}}
\newcommand{\rx}{{\rm x}}
\newcommand{\rs}{{\rm s}}
\newcommand{\rP}{{\rm P}}
\newcommand{\up}{\uparrow}
\newcommand{\down}{\downarrow}
\newcommand{\hc}{H_{\rm cond}}
\newcommand{\kb}{k_{\rm B}}
\newcommand{\cI}{{\cal I}}
\newcommand{\cM}{{\cal M}}
\newcommand{\cN}{{\cal N}}
\newcommand{\cm}{{m}}
\newcommand{\cn}{{n}}
\newcommand{\tit}{\tilde{t}}
\newcommand{\cE}{{\cal E}}
\newcommand{\cC}{{\cal C}}
\newcommand{\Ubs}{U_{\rm BS}}
\newcommand{\intO}{\int (dO)}
\sloppy

\title{Invariant integration over the orthogonal group} 
\author{Daniel Braun}
\affiliation{Laboratoire de Physique Th\'eorique, IRSAMC, UMR 5152 du CNRS,
  Universit\'e Paul Sabatier, 118, route de  
  Narbonne, 31062 Toulouse, FRANCE} 

\begin{abstract}
I adapt a recently introduced method for integrating over the unitary group
(S. Aubert and C.S. Lam, J.Math.Phys. 44, 6112-6131 (2003)) to the orthogonal
group. I derive 
explicit formulas for a number of 
one, two and three-vector integrals, as well as recursion formulas for more
complicated cases. 
\end{abstract}
\maketitle

\section{Introduction}
Integrals over compact Lie groups arise in physics naturally through
applications of random matrix theory \cite{Guhr98}, which plays an important
role in various fields 
of physics, ranging from nuclear theory \cite{Ullah64}, quantum chaos
\cite{Haake91} and  transport in mesoscopic devices \cite{Baranger94}, to
quantum 
information theory \cite{Zyczkowski03}. A new application arose recently in
the latter field 
from the study of statistics of interference in quantum algorithms
\cite{Braun06}. Quantum 
algorithms 
can always be represented as real matrices by spending one more qubit
(i.e. doubling the size of the Hilbert space), which labels the real or
imaginary part of the wave function \cite{Shi02,Aharonov03}, and the
question was posed how this 
influences the amount of interference necessary in the quantum algorithm
compared to a unitary representation \cite{Arnaud06}. Examining the
statistics of 
interference drawn from the Haar orthogonal ensemble, i.e.~the ensemble of
orthogonal matrices with a flat distribution according to the Haar measure
of the orthogonal group, then leads naturally to consider integrals of
monomials of matrix elements over the orthogonal group $O(N)$.

Formulas for one-vector integrals were obtained earlier by Ullah
\cite{Ullah64} and by Mehta \cite{Mehta91}. An $n-$vector integral is
defined as an integral containing 
elements from only $n$ columns (or from $n$ rows) of the orthogonal
matrix.
Prosen and co--workers introduced an asymptotic method which works well for
large $N$ \cite{Prosen02}. The problem was solved in principal by Gorin
\cite{Gorin02}, who derived 
recursion relations connecting $n-$vector integrals to $(n-1)-$vector
integrals. While the method is general, we
will see below that explicit formulas can be obtained in a much easier way
in various simple but important cases of integrals of low degree (and/or a
small number of vectors). A number of interesting properties of integrals
over the orthogonal (as well as the unitary and symplectic) group were also
derived in \cite{Collins04}, based on the use of Brauer algebras.

Recently, Aubert and Lam introduced a very elegant method for integration
over the unitary group $U(N)$ based solely on
the unitary constraint and the invariance of the Haar measure under
unitary transformations \cite{Aubert03}. This method is very powerful, and
allowed the authors to obtain 
explicit formulas for the lowest order integrals, including all one-vector
integrals, all two 
vector integrals with up to three different matrix elements taken to
arbitrary powers, and all two and
three-vector integrals up to order 6 (the order means here the number of
matrix elements in the monomial to be integrated). Recursion relations can
be 
obtained to reduce higher order integrals to these basic integrals. In the
present 
paper I adapt the method of invariant integration to the orthogonal group,
and derive  
the corresponding basic integrals and recursion formulae.

\section{Integrals over the orthogonal group}
Consider orthogonal, real $N\times N$ matrices $O$ with matrix elements
$O_{ij}$, $1\le i,j\le N$. We will be interested in integrals of the type
\begin{equation} \label{defI}
\cI_{i_1j_1\ldots i_pj_p}\equiv\int(dO)O_{i_1j_1}\ldots O_{i_pj_p}\,,
\end{equation}
where the order $p$ of the integral is some positive integer, and $(dO)$
denotes the Haar invariant measure of the orthogonal group $O(N)$,
normalized to $\int(dO)=1$. These integrals can be calculated based solely
on
\begin{itemize}
  \item the orthogonality relation $O^TO={\bf 1}=OO^T$ (where
  $^T$ denotes the transposed matrix), or explicitly
\begin{equation} \label{ortho}
\sum_{i=1}^NO_{ij}O_{ik}=\delta_{jk}=\sum_{i=1}^NO_{ji}O_{ki}\,,
\end{equation}
where $\delta_{jk}$ stands for the Kronecker-delta;
\item and the invariance of the Haar measure, meaning that for any  function
  $f(O)$ and an arbitrary (real) orthogonal ($N\times N$) matrix $V$, 
\begin{equation} \label{inv}
\int(dO)f(O)=\int(dO)f(VO)=\int(dO)f(OV)\,.
\end{equation}
Here we will use for $f(O)$ the monomial 
\begin{equation} \label{f}
f(O)=\prod_{\lambda=1}^p O_{i_\lambda j_\lambda}\equiv O_{IJ}\,,
\end{equation}
and we have introduced the notation $O_{IJ}$, where $I=\{i_1,\ldots,i_p\}$,
$J=\{j_1,\ldots,j_p\}$. Correspondingly, $\cI_{i_1j_1\ldots i_pj_p}$ will be
abbreviated as $\cI_{i_1j_1\ldots i_pj_p}\equiv \cI(O_{IJ})$. 
\end{itemize}

\section{Relations from invariance}
A number of powerful relations are obtained from choosing different explicit
orthogonal matrices, under which $\cI(O_{IJ})$ must be invariant.
\subsection{Global sign change}\label{gsc}
The simplest example consist of $V=-{\bf 1}$, i.e.~$V_{ij}=-\delta_{ij}$,
clearly an orthogonal transformation, with $O'=VO=-O$.  Eq.(\ref{inv})
implies $f(O)=(-1)^pf(O)$. Thus, $p$ needs to be even, otherwise the
integral will be zero. Even $p$ will therefore be assumed from now on.
\subsection{Local sign change}\label{lsc}
Another orthogonal transformation is induced by a matrix $V$ with matrix
elements $V_{ij}=(-1)^{s_i}\delta_{ij}$, where $s_i\in\{0,1\}$. It leads to
new matrix elements 
$O_{ij}'=(VO)_{ij}=(-1)^{s_i}O_{ij}$ and
$f(O')=(-1)^{\sum_{\lambda=1}^ps_{i_\lambda}}f(O)$. As each $s_{i_\lambda}$ is
arbitrary $\in\{0,1\}$, eq.(\ref{inv}) implies that each $i_\lambda$ has to
appear an even 
number of times. The same reasoning applied to multiplication with $V$
from the right lets us conclude that also each $j_\lambda$ has to appear an
even 
number of times, otherwise the integral will be zero.  
\subsection{Permutations}
Consider the permutation that exchanges two indices, $i_0\leftrightarrow
j_0$. It is induced by a transformation $V$ with matrix elements
$V_{ij}=\delta_{ij}$ for $i\ne i_0,j_0$, $V_{i_0j_0}=V_{j_0i_0}=1$,
$V_{i_0i_0}=V_{j_0j_0}=0$, which is manifestly orthogonal. An arbitrary
permutation can be obtained by concatenating exchanges of two indices. The
corresponding matrices are multiplied, and since the product of two
orthogonal matrices is again orthogonal, an arbitrary permutation $P$ can
always be represented as real orthogonal
matrix $V$.  Multiplication from the left permutes the left hand indices in
$O$, $f(VO)=\prod_\lambda O_{P(i_\lambda)j_\lambda}$, and multiplication from the right
permutes the right hand indices,  $f(OV)=\prod_\lambda O_{i_\lambda
  P(j_\lambda)}$. Therefore the value of the indices is of no importance, the
only thing which counts is the multiplicity of all different indices. Thus,
we may rewrite $I$ and $J$ as
$I=\{(i_1)^{\mu_1},(i_2)^{\mu_2},\ldots,(i_t)^{\mu_t}\}$ and
$J=\{(j_1)^{\nu_1},(j_2)^{\nu_2},\ldots,(j_s)^{\nu_s}\}$, with
$\sum_{i=1}^t\mu_i=p=\sum_{i=1}^s\nu_i$, where $p$ is an even integer according
to section \ref{gsc}, and all $\mu_i$ and $\nu_i$ must be even (section
\ref{lsc}). We might even drop the indices $i_\lambda$ and $j_\lambda$ all together
(e.g.~ chose them once for all as $i_\lambda=\lambda$ ($\lambda=1,\ldots,t$),
$j_\lambda=\lambda$ ($\lambda=1,\ldots,s$)), and just keep the multiplicities $\cM\equiv
\{\mu_1,\ldots,\mu_t\}$, $\cN\equiv\{\nu_1,\ldots,\nu_s\}$.  

\begin{figure}
\epsfig{file=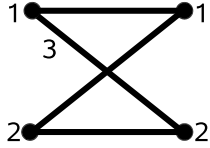,width=3cm,angle=0}
\caption{Example of a graphical representation for
  $\int(dO)O_{11}O_{12}^3O_{21}O_{22}$. Each dot in the left column presents
  a left index of a matrix element, each dot in the right column a right
  index of a matrix element. Each line corresponds to a factor in the
  monomial, with its power written next to it. The power is equal to one if
  it is not written out explicitly. The number of lines entering each point
  must be even.}\label{fig.graphs}        
\end{figure}

This result suggests a graphical representation of $\cI(O_{IJ})$, where all
first indices in $O_{IJ}$ are represented by a dot in a left column (one dot
for each different index), and all different second indices in $O_{IJ}$ are
represented by a dot in a right column. These dots are joined by lines,
where each line 
represents a factor $O_{i_\lambda j_\lambda}$. If a factor appears to the power
$m_\lambda$,
we denote that 
power next to the line (see 
fig. \ref{fig.graphs}). 
If all powers $m_\lambda$ are even, the integral is
positive, and is called a ``direct integral''; otherwise it is called
``exchange integral'' \cite{Aubert03}. In contrast to the integrals
encountered for the 
unitary group \cite{Aubert03}, there are no complex conjugate factors here.
This simplifies things at first, as we have to deal with only one type
of lines, but also complicates things as it will lead to larger
freedom in moving lines around (see below). Therefore, several recursion
relations and their solutions will turn out to be more complicated than in
the unitary case.

\subsection{Transposition}
We have $(dO)=(dO^T)$, and therefore $\cI(O_{IJ})=\cI(O_{JI})$. In the
graphical language this tells us that all diagrams are invariant
under reflection at a central vertical line.

\subsection{Rotations}\label{sec.rot}
Consider a rotation in the plane spanned by the basis vectors pertaining to
the indices $a$ and $b$, $V_{ij}=\delta_{ij}$ for $i\ne a,b$,
$V_{aa}=V_{bb}=\cos\xi$, $V_{ab}=-V_{ba}=(\sin\xi)$. Multiplied from the
right, this transforms matrix elements according to $O_{ia}'=(\cos\xi)
O_{ia}-(\sin\xi) O_{ib}$, $O_{ib}'=(\sin\xi)
O_{ia}+(\cos\xi) O_{ib}$, $O_{ij}'=O_{ij}$ otherwise. It is most instructive
to consider the effect of this rotation for a simple example, the integral
$\cI((O_{11})^d)$, $d$ even, and $a=1$, $b=2$. Expanding $(O_{11}')^d$, we
obtain 
\begin{equation} \label{IO11}
\cI((O_{11}')^d)=\sum_{e=0}^d{d\choose
e}(\cos\xi)^{d-e}(-\sin\xi)^{e} \int
(dO)(O_{11})^{d-e}(O_{12})^e=\cI((O_{11})^d)\,,
\end{equation}
where the last equal sign is dictated by the invariance under rotations, so
the left hand side must be independent of $\xi$. As the only independent
form of $\cos\xi$ and of $\sin\xi$ are powers of $\cos^2\xi+\sin^2\xi$, the
left side can only contain even values of $e$,  such that with
$e=2\tilde{e}$, $d=2\tilde{d}$
\begin{eqnarray}
\cI((O_{11}')^d)&=&\sum_{\tilde{e}=0}^{\tilde{d}}{2\tilde{d}\choose
2\tilde{e}}(\cos\xi)^{2(\tilde{d}-\tilde{e})}(\sin\xi)^{2\tilde{e}}\cI((O_{11})^{2(\tilde{d}-\tilde{e})}(O_{12})^{2\tilde{e}})\\
&=&\sum_{\tilde{e}
 =0}^{\tilde{d}}M_e(\cos\xi)^{2(\tilde{d}-\tilde{e})}(\sin\xi)^{2\tilde{e}}\,,\label{expan}\\  
M_e&=&  {2\tilde{d}\choose
2\tilde{e}}\cI((O_{11})^{2(\tilde{d}-\tilde{e})}(O_{12})^{2\tilde{e}})\,.\label{Me}
\end{eqnarray}
Invariance under rotations requires that $M_e\propto {\tilde{d}\choose
\tilde{e}}$. The proportionality constant can be fixed from
$\tilde{e}=0$, $M_0=\cI((O_{11})^d)\equiv F_1(2\tilde{d})$, and we are thus
lead to $M_e={\tilde{d}\choose 
\tilde{e}}\cI((O_{11})^d)$. Together with (\ref{Me}) this leads to a
series of relationships,
\begin{equation} \label{F2}
F_2(2(\tilde{d}-\tilde{e}),2\tilde{e})\equiv
\cI((O_{11})^{2(\tilde{d}-\tilde{e})}(O_{12})^{2\tilde{e}})=\frac{ {\tilde{d}\choose
\tilde{e}}}{{2\tilde{d}\choose
2\tilde{e}}}\cI((O_{11})^d)=\frac{ {\tilde{d}\choose
\tilde{e}}}{{2\tilde{d}\choose
2\tilde{e}}}F_1(2\tilde{d})\,.
\end{equation}
In the graphical language, we have rotated $2\tilde{e}$ lines away from the
right dot to a new, empty dot, and we have found a relationship between this
new integral and the old one, containing only one line of multiplicity
$d=2\tilde{d}$. The same reasoning can be applied to a dot which is part of
a more complicated diagram, i.e.~a dot in which lines from several other
dots terminate. The total number of lines arriving at the dot will again be
denoted by $d$, and must still be even. Invariance under rotations implies
again the same form (\ref{expan}) after expanding the powers of all matrix
elements which get transformed, and leads to the same eq.(\ref{F2}), where
now $F_1(2\tilde{d})$ means the original diagram, and
$F_2(2(\tilde{d}-\tilde{e}),2\tilde{e})$ the diagram in which $2\tilde{e}$
lines have been rotated away from one dot to another, empty dot. Note that
exactly the same relation is obtained for integration over the hypersphere
(instead of integration over the orthogonal group) \cite{Aubert03}. This is
not surprising, as (\ref{IO11}) is a one--vector integral, such that only
the normalization of each column is involved, but no orthogonality relation
between different columns. More generally, any one--vector integral
over the orthogonal group $O(N)$ equals the corresponding integral over the
hypersphere in $N$ dimensions \cite{Gorin02}.

\begin{figure}
\epsfig{file=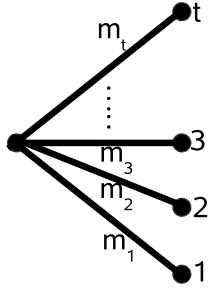,width=3cm,angle=0}
\caption{Graphical representation of the fan integral,
  eq.(\ref{Fn2}).}\label{fig.fan}         
\end{figure}

Eq.(\ref{F2}) can be iterated to give the relations called ``fan relations''
in \cite{Aubert03}, 
\begin{equation} \label{Fn}
F_t(2(\tilde{d}-\tilde{d}_1),2(\tilde{d}_1-\tilde{d}_2),\ldots,2(\tilde{d}_{t-2}-\tilde{d}_{t-1}),2\tilde{d}_{t-1}))=
F_1(2\tilde{d})\frac{{\tilde{d}\choose\tilde{d}_1}{\tilde{d_1}\choose\tilde{d}_2}\cdot\ldots\cdot{\tilde{d}_{t-2}\choose\tilde{d}_{t-1}}}{{2\tilde{d}\choose2\tilde{d}_1}{2\tilde{d}_1\choose2\tilde{d}_2}\cdot\ldots\cdot{2\tilde{d}_{t-2}\choose2\tilde{d}_{t-1}}}  \,.
\end{equation}
We recognize $2\tilde{d}_{t-1}=m_t$ as the multiplicity of the last line (line
$t$), $2(\tilde{d}_{t-2}-\tilde{d}_{t-1})=m_{t-1}$ as the multiplicity of
line $t-1$, and so on, up to the multiplicity $m_1=2(\tilde{d}-\tilde{d}_1)$
of the first line. The relationships between $\tilde{d}_i$ and $m_j$ are
easily inverted, and, when re-injected into (\ref{Fn}), lead to the final
form of the ``fan relation'' for an integral represented by the diagram in
fig.\ref{fig.fan}, 
\begin{equation} \label{Fn2}
F_t(m_1,\ldots,m_t)=F_1(d)\frac{(d/2)!}{d!}\frac{m_1!\cdot\ldots\cdot m_t!}{(\frac{m_1}{2})!\cdot\ldots\cdot(\frac{m_t}{2})!}\,.
\end{equation}
The integral for a single line with multiplicity $d$, $F_1(d)$, will be
calculated in the next section.

\section{Relations from Orthogonality}
The orthogonality relation (\ref{ortho}) is a handy tool to reduce the
number of lines in a diagram. It is most instructive to consider as
simple example the integral 
\begin{equation} \label{f2}
F_2(2(m-1),2)=\intO (O_{1k})^{2(m-1)}(O_{1l})^2\,\,\,\,\,k\ne l\,,
\end{equation}
As we have seen, the value of
the integral is independent of the index $l$, as long as $l\ne k$. At the
same time, if we sum over {\em all} possible values of  $l$
(i.e.~$l=1,\ldots,N$), we can make use of (\ref{ortho}), as
$\sum_{l=1}^N(O_{1l})^2=1$. Therefore, 
\begin{eqnarray} \label{F21}
(N-1)F_2(2(m-1),2)&=&\sum_{l=1}^N\intO(O_{1k})^{2(m-1)}(O_{1l})^2-\intO(O_{1k})^{2m}\nonumber\\
&=&F_1(2(m-1))-F_1(2m)\,.
\end{eqnarray}
This is an equation independent of the fan relation (\ref{Fn2}). The latter
leads in the present case to 
\begin{equation} \label{F22}
F_2(2(m-1),2)=\frac{F_1(2m)}{2m-1}\,,
\end{equation}
and combining this with (\ref{F21}) gives a recursion relation for $F_1(2m)$, 
\begin{equation} \label{recurF}
F_1(2m)=\frac{2m-1}{2m+N-2}F_1(2m-2)\,,
\end{equation}
with the obvious solution
\begin{equation} \label{F2m}
F_1(2m)=\frac{(2m-1)!!(N-2)!!}{(2m+N-2)!!}\,.
\end{equation}
The normalization $F_1(0)=\intO=1$ was used, and can be retrieved from
(\ref{F2m}) 
if we define $(-1)!!=1$.

The orthogonality relation is  useful if we have a dot in which only two
lines end. Note that so far we have used again just the
normalization of each line in the orthogonal matrix. If the two lines ending
in dot $l$ on the right were replaced by two lines originating from two
different dots on the left, the sum over all values of $l$ would give zero,
regardless of the rest of the diagram. 

\section{Z--integrals}
\begin{figure}
\epsfig{file=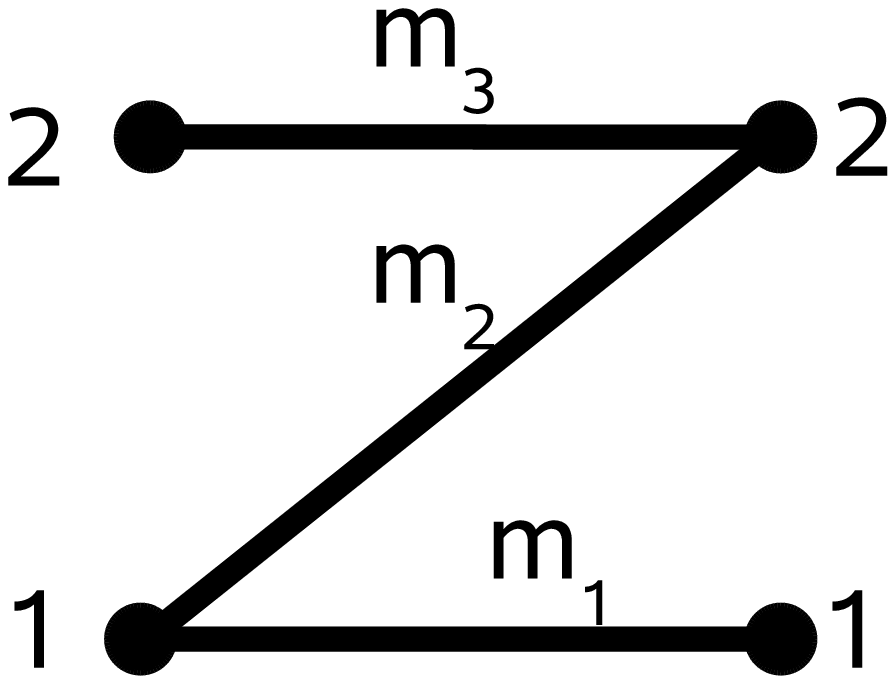,width=3cm,angle=0}
\epsfig{file=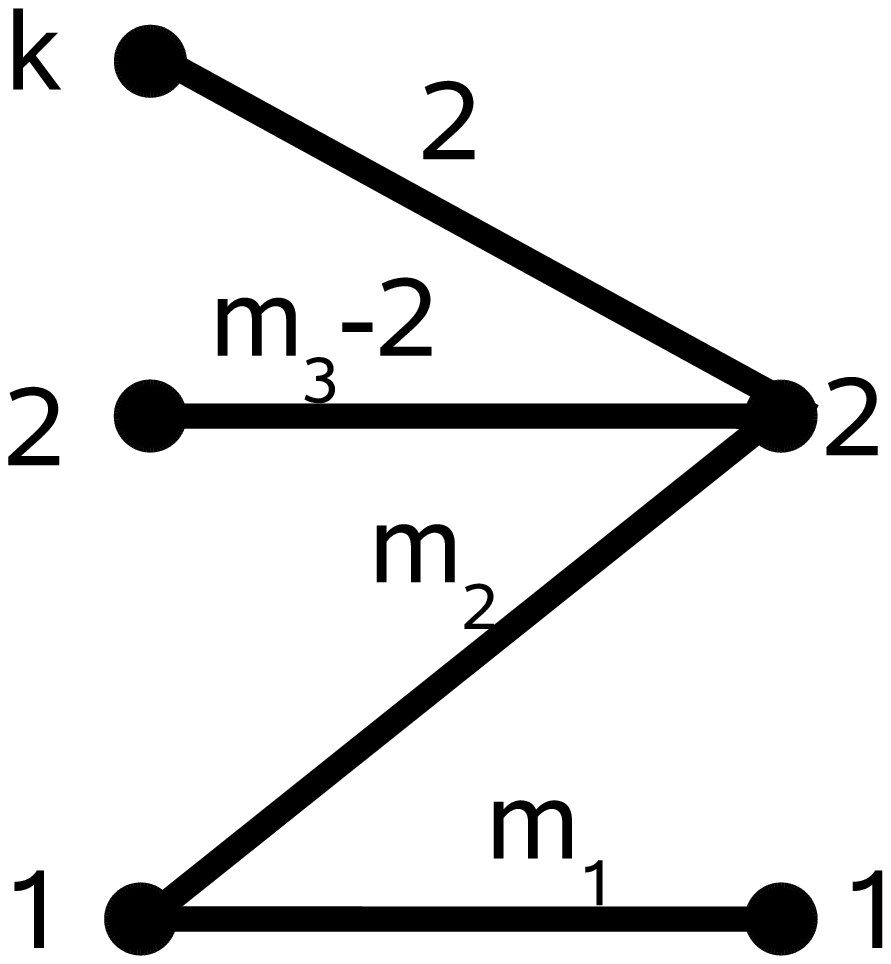,width=3cm,angle=0}
\caption{Graphical representation of the Z--integral,
  eq.(\ref{Z}) (left) and the integral ${\cal I}(3b)$ used to establish a
  recursion 
  relation for it (right).}\label{fig.Z}         
\end{figure}

As an application of the techniques introduced, we now consider all possible
integrals with up to three different factors, two indices on the right,
and two on the left (see fig.\ref{fig.Z}), baptized ``Z--integrals'' in
\cite{Aubert03},
\begin{equation} \label{Z}
Z(m_1,m_2,m_3)=\intO (O_{11})^{m_1}(O_{12})^{m_2}(O_{22})^{m_3}\,.
\end{equation}
All multiplicities $m_1,m_2,m_3$ must be even. We start from the integral
$\cI(3b)$ shown in fig.\ref{fig.Z}, obtained by rotating two lines away from
the first dot on the left to a new dot. The latter is arbitrary, and we can
sum over it, avoiding the already taken indices 1 and 2, whereas in the full
sum the extra line disappears,
\begin{equation} \label{sum3b}
\sum_{k=3}^N\cI(3b)=(N-2)\cI(3b)=Z(m_1,m_2,m_3-2)-Z(m_1,m_2,m_3)-Z(m_1,m_2+2,m_3-2)\,. 
\end{equation}
On the other hand, the fan relation (\ref{Fn2}) can be applied to the upper
two lines of the diagram (3b), which gives
$\cI(3b)=\frac{1}{m_3-1}Z(m_1,m_2,m_3)$. Once we insert the latter result
into (\ref{sum3b}), 
we find the recursion
\begin{equation} \label{recurZ}
Z(m_1,m_2,m_3)=\frac{m_3-1}{N+m_3-3}\left(Z(m_1,m_2,m_3-2)-Z(m_1,m_2+2,m_3-2)\right)\,, 
\end{equation}
which should be supplemented by the initial value
\begin{equation} \label{Zanf}
Z(m_1,m_2,0)=F_2(m_1,m_2)=\frac{(m_1+m_2-1)!!(N-2)!!}{(m_1+m_2+N-2)!!}\frac{(\frac{m_1+m_2}{2})!m_1!m_2!}{(m_1+m_2)!(\frac{m_1}{2})!(\frac{m_2}{2})!}
\end{equation}
obtained from the fan relation (\ref{Fn2}). If we apply the recursion
relation again, we easily convince ourselves that its solution is of the
form 
\begin{equation} \label{Z1}
Z(m_1,m_2,m_3)=\frac{(m_3-1)!!(N-3)!!}{(N+m_3-3)!!}
 \sum_{i=0}^{m_3/2}{m_3/2\choose  i}Z(m_1,m_2+2i,0)(-1)^i\,.  
\end{equation}
A closed form can be found for the sum in this equation, if we
consider $N$ even and odd separately. For $N$ even, (\ref{Zanf}) can be
re-expressed as
\begin{equation} \label{Zanf2}
Z(m_1,m_2,0)=\frac{\left(\frac{N-2}{2}\right)!m_1!m_2!}{2^{m_1+m_2}(\frac{m_1}{2})!(\frac{m_2}{2})!(\frac{m_1+m_2+N-2}{2})!}\,,
\end{equation}
such that 
\begin{eqnarray}
Z(m_1,m_2,m_3)&=&\frac{(m_3-1)!!(N-3)!!}{(N+m_3-3)!!}
\frac{\left(\frac{N-2}{2}\right)!m_1!}{2^{m_1+m_2}(\frac{m_1}{2})!} 
S_1\,,\nonumber\\
S_1&\equiv&\sum_{i=0}^{m_3/2}\left(\frac{-1}{4}\right)^i{m_3/2\choose
  i}\frac{(m_2+2i)!}{(\frac{m_2}{2}+i)!(\frac{m_1+m_2+2i+N-2}{2})!}\nonumber\\
&=&\frac{2^\frac{m_2-m_3}{2}(m_2-1)!!(m_1+m_3+N-3)!!}{(m_1+N-3)!!(\frac{m_1+m_2+m_3+N-2}{2})!}\,.\label{Zfin}
\end{eqnarray}
This leads to the final result
\begin{eqnarray}
Z(m_1,m_2,m_3)&=&\frac{m_1!m_2!m_3!\left(\frac{m_1+N-2}{2}\right)!\left(\frac{m_3+N-2}{2}\right)!}{2^{m_1+m_2+m_3}(m_1+N-2)!(m_3+N-2)!\left(\frac{m_1+m_3+N-2}{2}\right)!\left(\frac{m_1+m_2+m_3+N-2}{2}\right)!}\nonumber\\
&&\times\frac{(m_1+m_3+N-2)!(N-2)!}{\left(\frac{m_1}{2}\right)!\left(\frac{m_2}{2}\right)!\left(\frac{m_3}{2}\right)!}
\\
&=&\frac{2^{2-N}\Gamma(\frac{1+m_1}{2})\Gamma(\frac{1+m_2}{2})\Gamma(\frac{1+m_3}{2})\Gamma(N-1)\Gamma(\frac{1}{2}(N+m_1+m_3-1))}{\pi\Gamma(\frac{1}{2}(N+m_1-1))\Gamma(\frac{1}{2}(N+m_3-1))\Gamma(\frac{1}{2}(N+m_1+m_2+m_3))}\,.
\end{eqnarray}
The latter form turns out to be valid also for odd $N$.

\section{Exchange integrals}
\begin{figure}
\epsfig{file=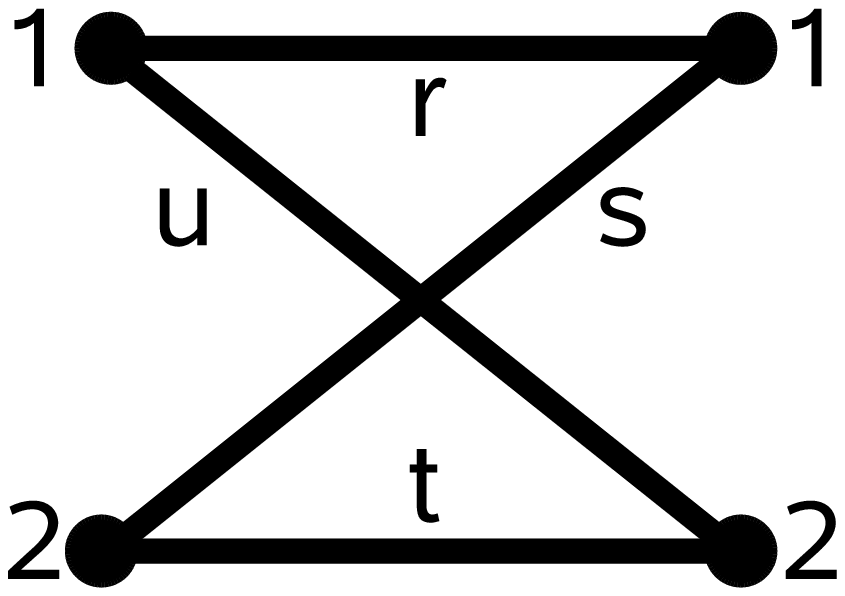,width=3cm,angle=0}
\epsfig{file=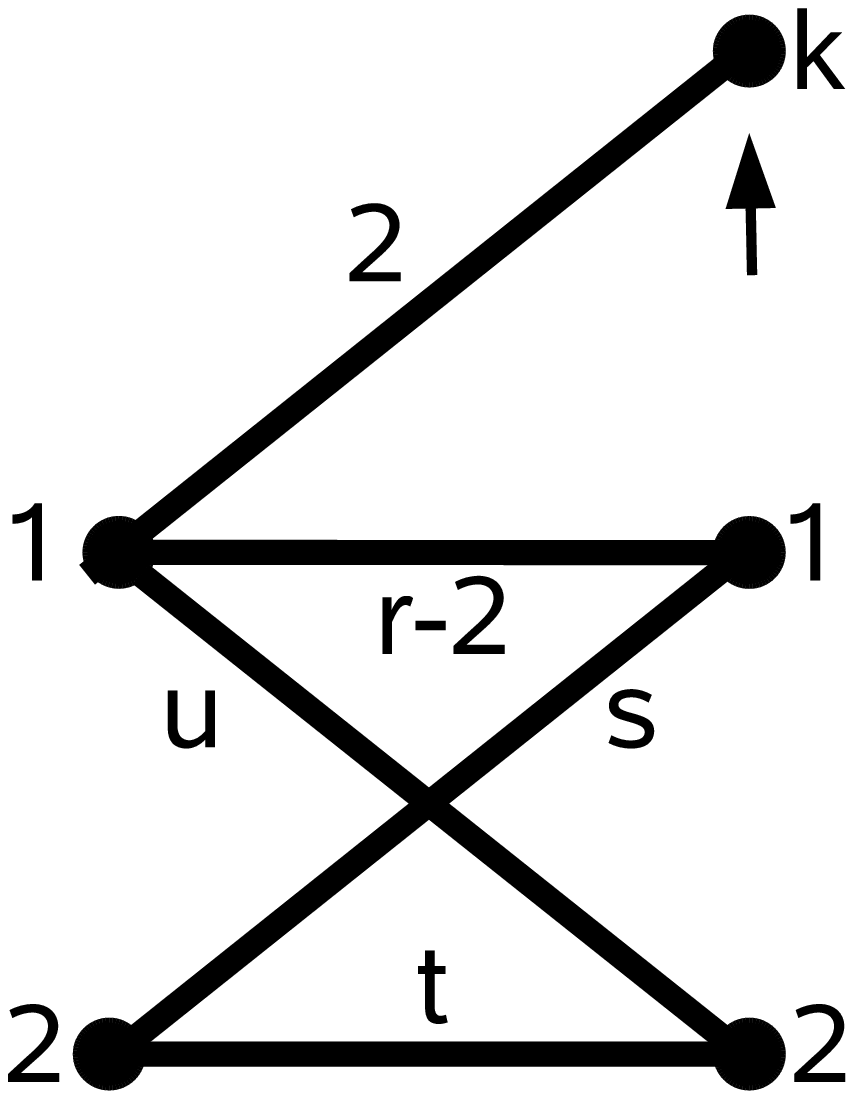,width=3cm,angle=0}
\epsfig{file=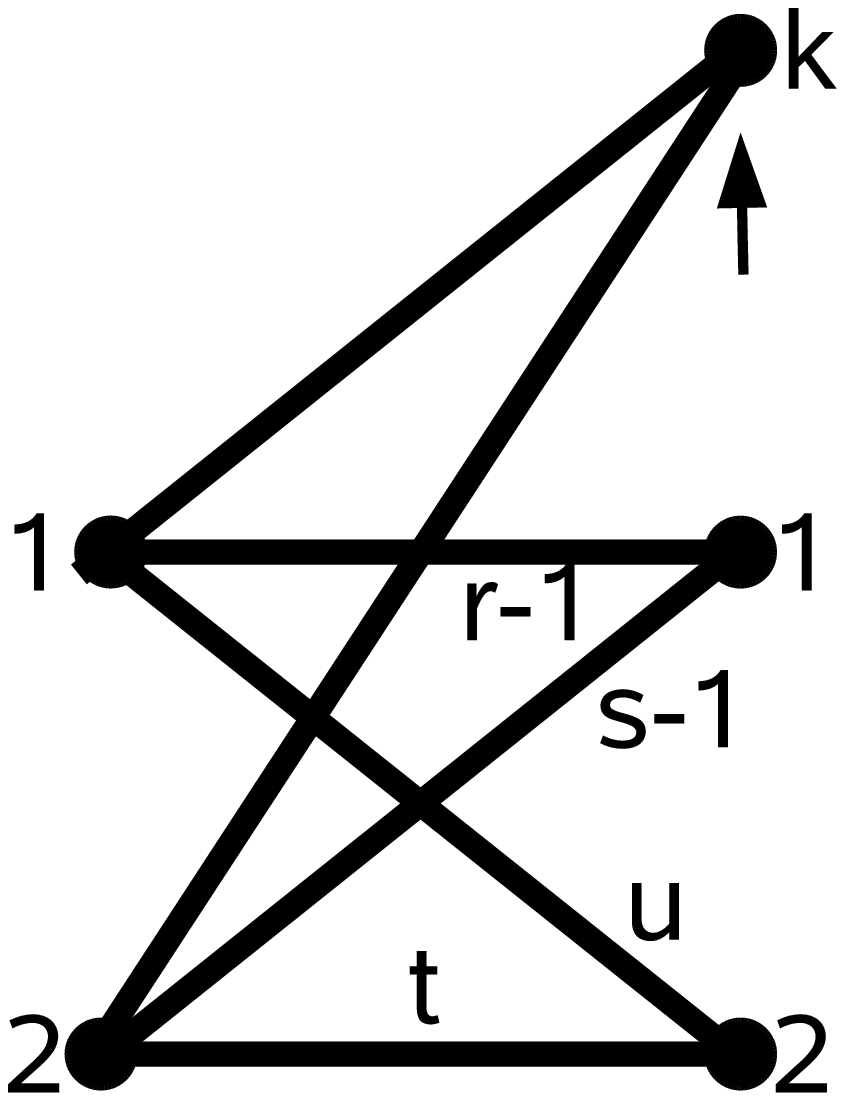,width=3cm,angle=0}
\epsfig{file=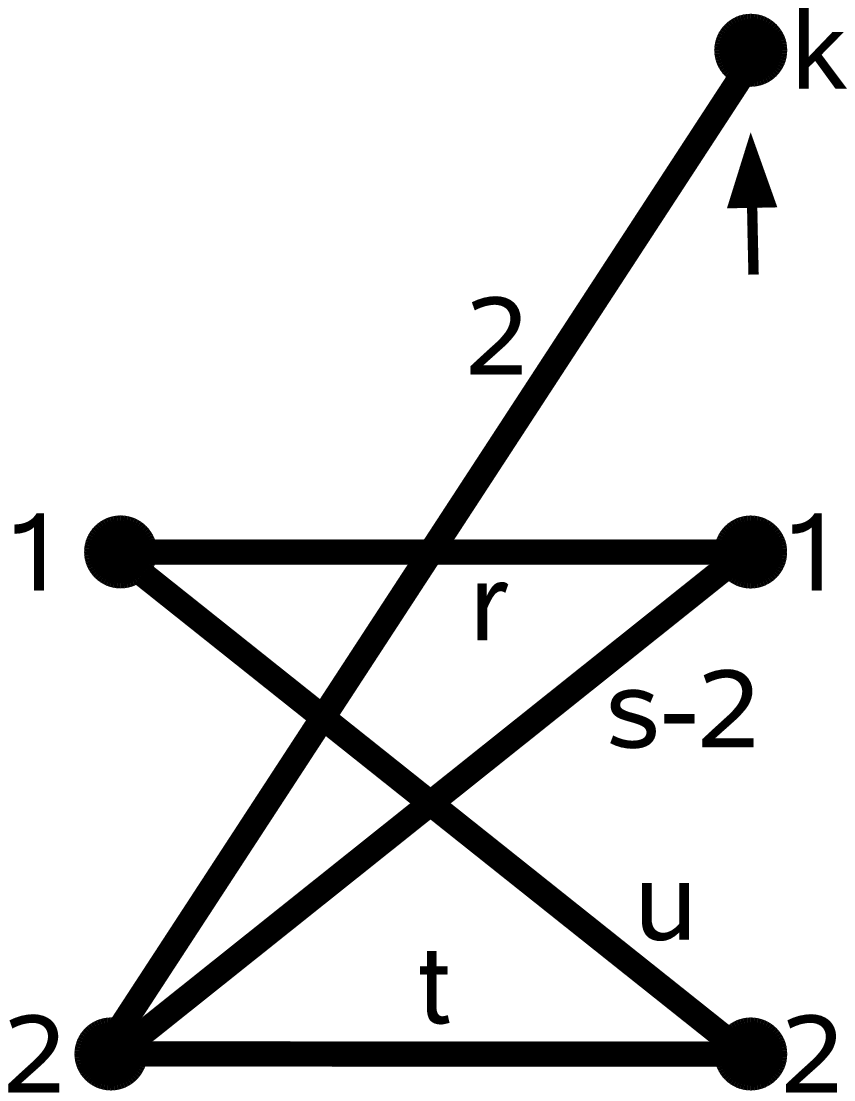,width=3cm,angle=0}
\caption{The exchange integral $X(r,s,t,u)$ (left) and the integrals
  $\cI(4a)$, $\cI(4b)$, and $\cI(4c)$ used to calculate it (second from left
  to right, respectively). The latter integrals are obtained by rotating 
  two lines away from dot 1 on the right to a new dot $k\ne 1,2$; the index
  $k$ is then summed over.} \label{fig.X}         
\end{figure}

Exchange integrals contain at least one line with odd multiplicity, and can
therefore be positive or negative. The total number of lines arriving in any
dot must, of course, still be even, otherwise the integral
vanishes. Consider the structurally simplest exchange integral depicted in
fig.\ref{fig.X}, 
\begin{equation} \label{X}
X(r,s,t,u)\equiv \intO (O_{11})^r(O_{21})^s(O_{22})^t(O_{12})^u\,,
\end{equation}
$r,s,t,u\in \mathbb{N}$. The case $r=s=t=u=1$ is easily solved by summing
over the index of an arbitrary point, say the upper right one.
We obtain 
\begin{equation} \label{X1111}
X(1,1,1,1)=-\frac{1}{N-1}F_2(2,2)=-\frac{1}{(N-1)N(N+2)}\,.
\end{equation}
For general $r,s,t,u$ we derive a recursion relation
using invariance under a rotation between indices 1 and 3 (where 3 is a new
index), multiplied from the right. This gives
\begin{eqnarray}
X(r,s,t,u)=X'(r,s,t,u)&\equiv&\intO\left((\cos\xi) O_{11}-(\sin\xi)
O_{13}\right)^r\left((\cos\xi) O_{21}-(\sin\xi)
O_{23}\right)^sO_{22}^tO_{12}^u\nonumber\\
&=&(\cos\xi)^{r+s}M_0+(\cos\xi)^{r+s-2}(\sin\xi)^2 M_1+\ldots
\end{eqnarray}
where $M_0=X(r,s,t,u)$ is evident from choosing $\xi=0$, and $M_1$ is given
by the diagrams in fig.\ref{fig.X} with corresponding values $\cI(4a)$,
$\cI(4b)$, and $\cI(4c)$,
\begin{equation} \label{M1X}
M_1=\frac{r(r-1)}{2}\cI(4a)+rs\cI(4b)+\frac{s(s-1)}{2}\cI(4c)\,.
\end{equation}
According to section \ref{sec.rot}, we must have
$M_1=\frac{r+2}{2}X(r,s,t,u)$. 
The order of the integrals $\cI(4a)$,
$\cI(4b)$, and $\cI(4c)$ can again be
reduced by summing over the index of the new point, which yields
\begin{eqnarray}
\cI(4a)&=&\frac{1}{N-2}\left(X(r-2,s,t,u)-X(r,s,t,u)-X(r-2,s+2,t,u)\right)\\
\cI(4b)&=&-\frac{1}{N-2}\left(X(r,s,t,u)+X(r-1,s-1,t+1,u+1)\right)\\
\cI(4c)&=&\frac{1}{N-2}\left(X(r,s-2,t,u)-X(r,s-2,t+2,u)-X(r,s,t,u)\right)\,.
\end{eqnarray}
If we insert these integrals into (\ref{M1X}), we get the recursion relation
\begin{eqnarray}
X(r,s,t,u)&=&\frac{1}{(r+s)(N-2)+r(r-1)+2rs+s(s-1)}\big(-2rsX(r-1,s-1,t+1,u+1)\nonumber\\ 
&&+r(r-1)\left(X(r-2,s,t,u)-X(r-2,s+2,t,u)\right)\nonumber\\
&&+s(s-1)\left(X(r,s-2,t,u)-X(r,s-2,t+2,u)\right)\big)\,.\label{Xrec}
\end{eqnarray}
No closed solution of this recursion relation could be found, but the
principle of its application is clear: Iterating (\ref{Xrec}) will reduce
the multiplicities of the lines with power $r$ or $s$, till a power zero is
achieved, and then the integral is reduced to a Z-integral, whose value is
known for all multiplicities of the remaining lines, see eq.(\ref{Zfin}).
Note that the formula corresponding to (\ref{Xrec}) for the unitary case is
quite different, 
as one must rotate always one line corresponding to a
non-conjugated matrix 
element and  one line corresponding to a complex conjugated matrix element
\cite{Aubert03}. 

\section{Integrals of order 6}
\begin{figure}
\epsfig{file=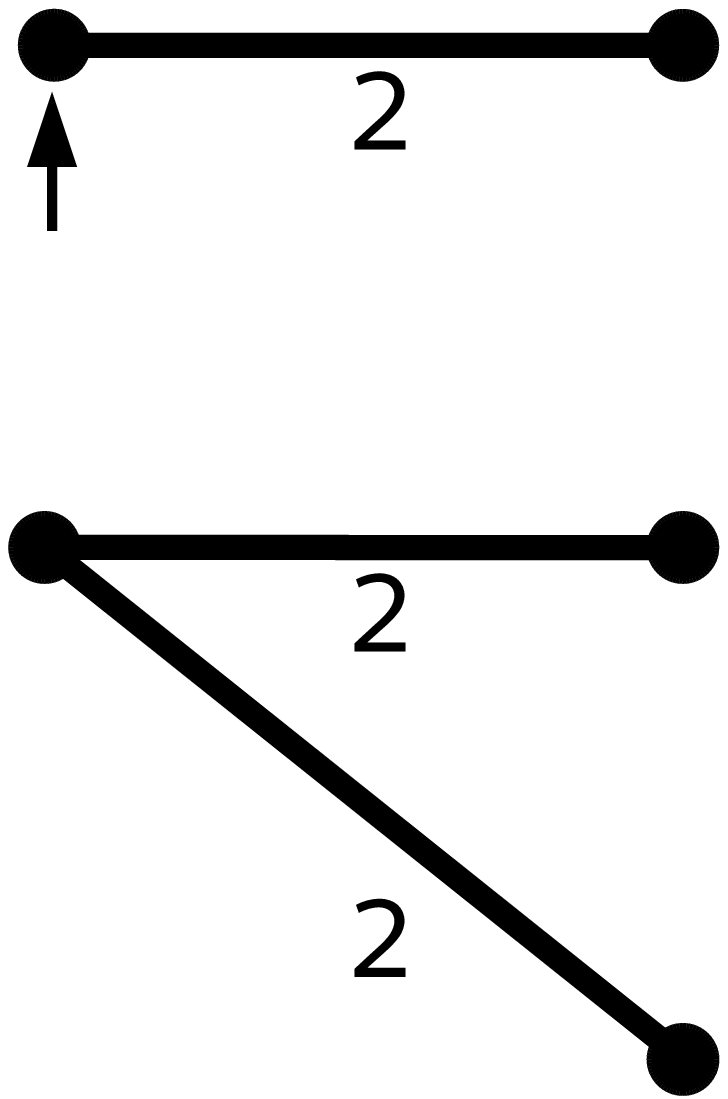,width=2cm,angle=0}
\epsfig{file=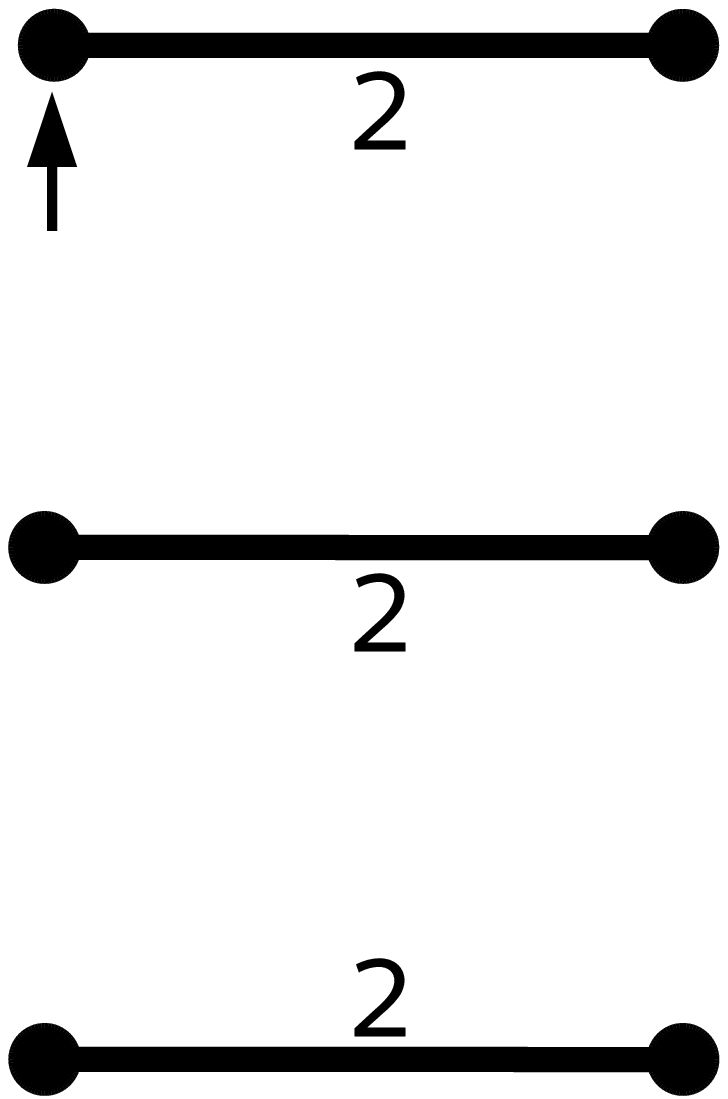,width=2cm,angle=0}
\epsfig{file=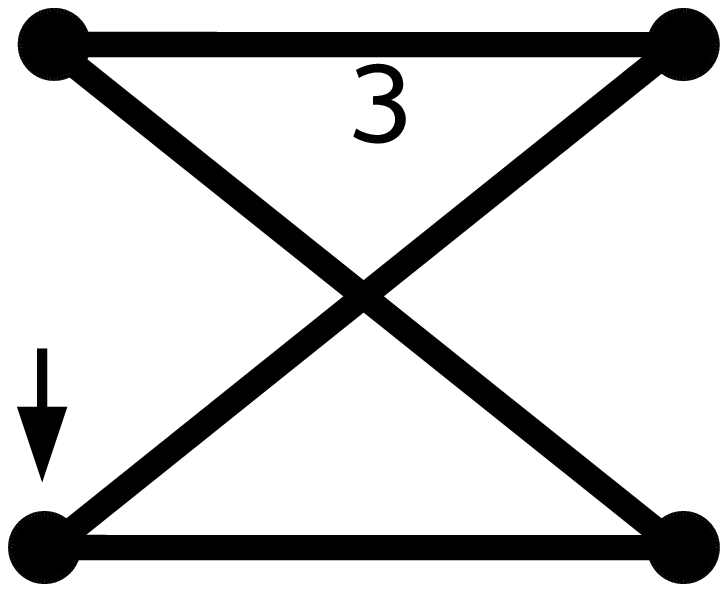,width=2cm,angle=0}
\epsfig{file=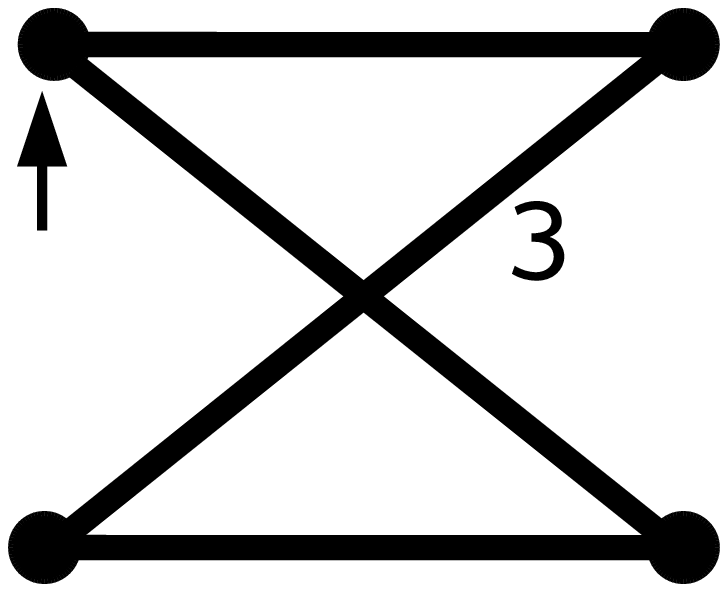,width=2cm,angle=0}
\epsfig{file=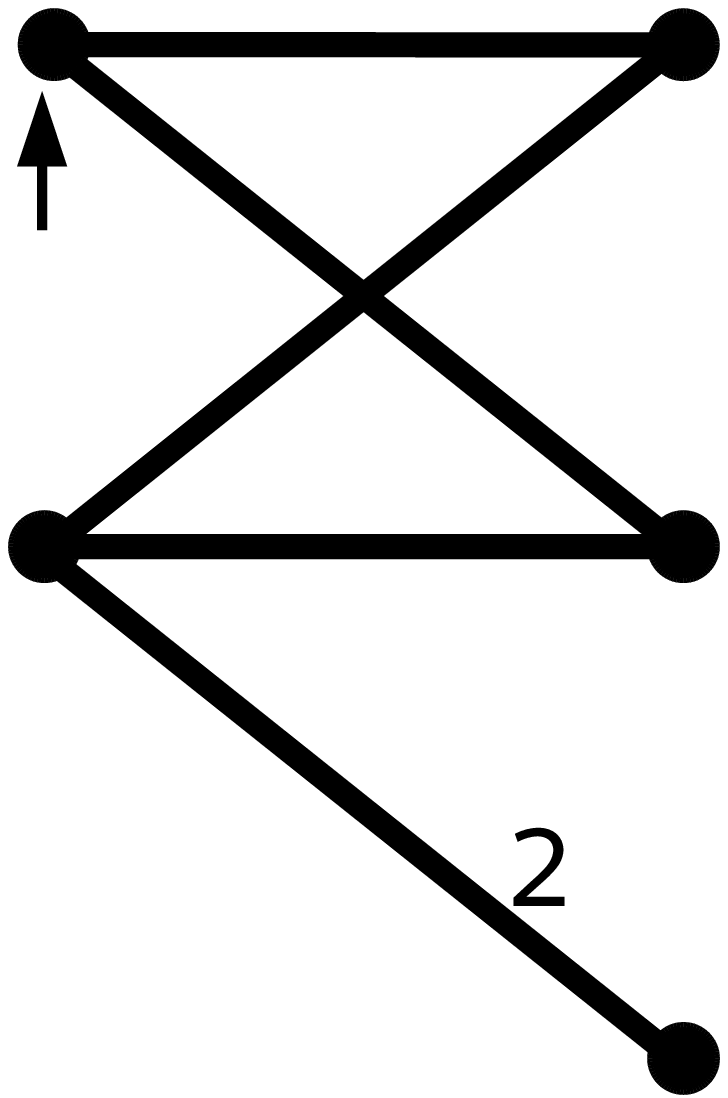,width=2cm,angle=0}
\epsfig{file=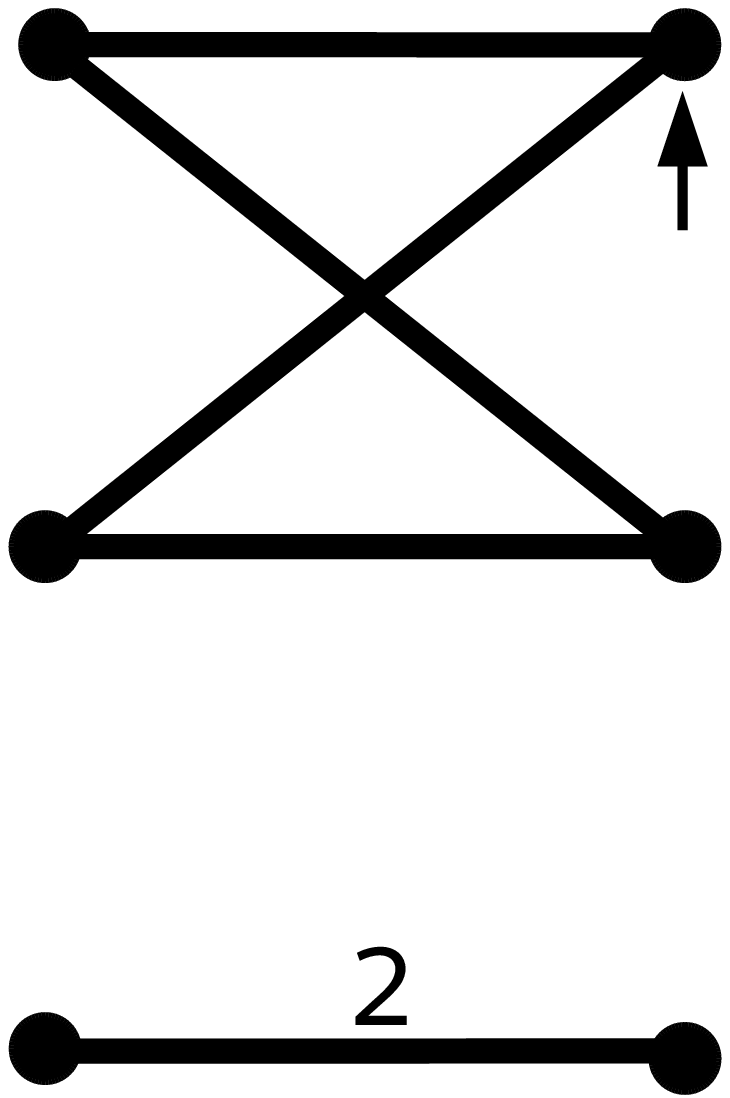,width=2cm,angle=0}
\epsfig{file=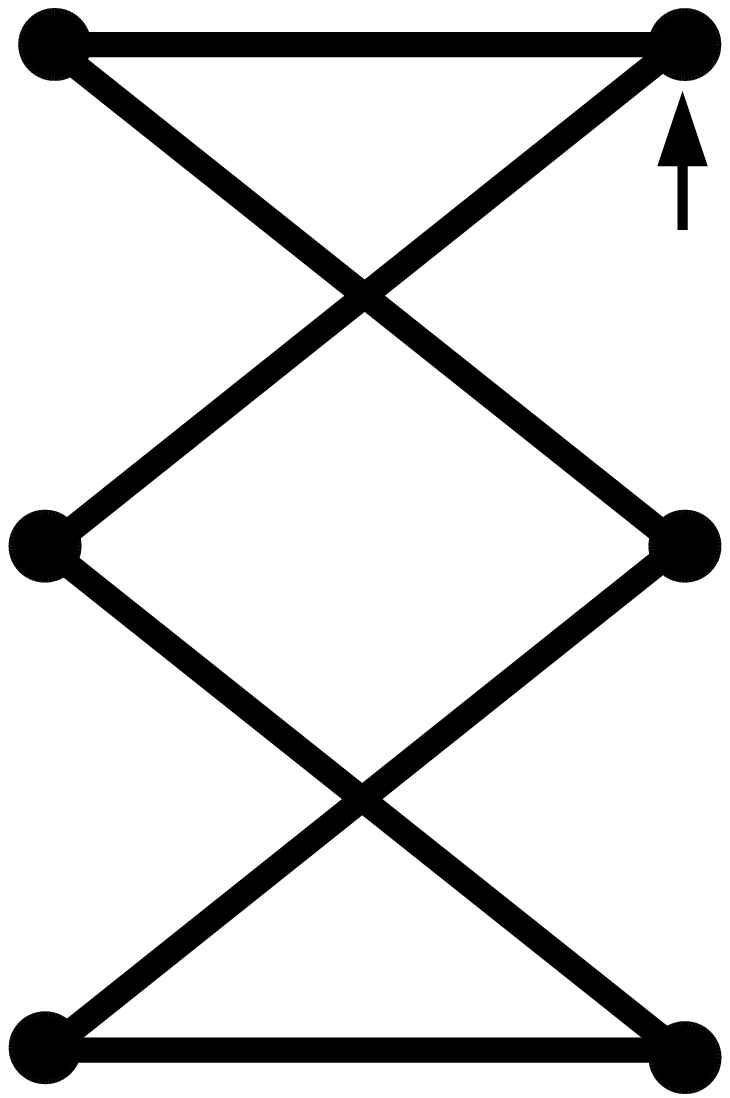,width=2cm,angle=0}
\caption{All the integrals of degree $p=6$ which are not of the fan-type
or Z-type considered previously. These integrals are
  labeled $\cI(5a)$, $\cI(5b)$, $\cI(5c)$, $\cI(5d)$, $\cI(5e)$, $\cI(5f)$,
  $\cI(5g)$ (from left to right), respectively, and can
  be obtained by summing over the indices of 
  the dots with an arrow.  $\cI(5c)$ and $\cI(5d)$ are of the exchange type
  considered in fig.\ref{fig.X}.} \label{fig.p6}         
\end{figure}

In this final section I will give explicit formulas for all integrals of
order $p=6$, which are not of the fan-type or Z-type. They are shown in
figure \ref{fig.p6}, and will be denoted 
accordingly $\cI(5a)$, $\cI(5b)$, $\cI(5c)$, $\cI(5d)$, $\cI(5e)$,
$\cI(5f)$, $\cI(5g)$. All of them are obtained by summing over the index
indicated by an arrow. We are lead to  
\begin{eqnarray}
\cI(5a)&=&\frac{1}{N-1}\left(F_2(2,2)-F_3(2,2,2)\right)=
\frac{N+3}{(N-1)N(N+2)(N+4)}\\    
\cI(5b)&=&\frac{1}{N-2}(Z(2,0,2)-2\cI(5a))=
\frac{-2+N(N+3)}{(N-2)(N-1)N(N+2)(N+4)} \\ 
\cI(5c)&=&-\frac{1}{N-1}F_2(2,4)=-\frac{3(N-2)!!}{(N-1)(N+4)!!}=\cI(5d)\\
\cI(5e)&=&-\frac{1}{N-1}F_3(2,2,2)=-\frac{(N-2)!!}{(N-1)(N+4)!!}\\
\cI(5f)&=&-\frac{1}{N-2}\left(\cI(5a)+\cI(5e)\right)=-\frac{1}{(N-2)(N-1)N(N+4)}\\ 
\cI(5g)&=&-\frac{2}{N-2}\cI(5e)=\frac{2(N-2)!!}{(N-2)(N-1)(N+4)!!}\,.
\end{eqnarray}

\section{Summary}
As a summary, I have adapted the method of invariant integration introduced
in \cite{Aubert03} to the case of integration over the orthogonal
group $O(N)$. Explicit formulas were obtained for all one--vector integrals
(these 
coincide with integrals over a hypersphere in $N$ dimensions), for all
two--vector integrals with up to three different matrix elements, as well as
for all integrals up to order 6. In more complicated cases recursion
relations were derived, in particular for all exchange integrals with two
different indices on the left and two on the right.

{\em Acknowledgments:}
This work
was supported by the Agence 
National de la Recherche 
(ANR), project INFOSYSQQ.
\bibliography{../mybibs_bt}

\begin{thebibliography}{14}
\expandafter\ifx\csname natexlab\endcsname\relax\def\natexlab#1{#1}\fi
\expandafter\ifx\csname bibnamefont\endcsname\relax
  \def\bibnamefont#1{#1}\fi
\expandafter\ifx\csname bibfnamefont\endcsname\relax
  \def\bibfnamefont#1{#1}\fi
\expandafter\ifx\csname citenamefont\endcsname\relax
  \def\citenamefont#1{#1}\fi
\expandafter\ifx\csname url\endcsname\relax
  \def\url#1{\texttt{#1}}\fi
\expandafter\ifx\csname urlprefix\endcsname\relax\def\urlprefix{URL }\fi
\providecommand{\bibinfo}[2]{#2}
\providecommand{\eprint}[2][]{\url{#2}}

\bibitem[{\citenamefont{Guhr et~al.}(1998)\citenamefont{Guhr,
  M\"uller-Gr\"oling, and Weidenm\"uller}}]{Guhr98}
\bibinfo{author}{\bibfnamefont{T.}~\bibnamefont{Guhr}},
  \bibinfo{author}{\bibfnamefont{A.}~\bibnamefont{M\"uller-Gr\"oling}},
  \bibnamefont{and} \bibinfo{author}{\bibfnamefont{H.~A.}
  \bibnamefont{Weidenm\"uller}}, \bibinfo{journal}{Phys. Rep.}
  \textbf{\bibinfo{volume}{299}}, \bibinfo{pages}{190} (\bibinfo{year}{1998}).

\bibitem[{\citenamefont{Ullah}(1964)}]{Ullah64}
\bibinfo{author}{\bibfnamefont{N.}~\bibnamefont{Ullah}},
  \bibinfo{journal}{Nucl. Phys.} \textbf{\bibinfo{volume}{58}},
  \bibinfo{pages}{65} (\bibinfo{year}{1964}).

\bibitem[{\citenamefont{Haake}(1991)}]{Haake91}
\bibinfo{author}{\bibfnamefont{F.}~\bibnamefont{Haake}},
  \emph{\bibinfo{title}{{Quantum Signatures of Chaos}}}
  (\bibinfo{publisher}{Springer}, \bibinfo{address}{Berlin},
  \bibinfo{year}{1991}).

\bibitem[{\citenamefont{Baranger and Mello}(1994)}]{Baranger94}
\bibinfo{author}{\bibfnamefont{H.~U.} \bibnamefont{Baranger}} \bibnamefont{and}
  \bibinfo{author}{\bibfnamefont{P.~A.} \bibnamefont{Mello}},
  \bibinfo{journal}{Phys. Rev. Lett.} \textbf{\bibinfo{volume}{73}},
  \bibinfo{pages}{142} (\bibinfo{year}{1994}).

\bibitem[{\citenamefont{\.{Z}yczkowski and Sommers}(2003)}]{Zyczkowski03}
\bibinfo{author}{\bibfnamefont{K.}~\bibnamefont{\.{Z}yczkowski}}
  \bibnamefont{and} \bibinfo{author}{\bibfnamefont{H.-J.}
  \bibnamefont{Sommers}}, \bibinfo{journal}{J. Phys. A}
  \textbf{\bibinfo{volume}{36}}, \bibinfo{pages}{10115} (\bibinfo{year}{2003}).

\bibitem[{\citenamefont{Braun and Georgeot}(2006)}]{Braun06}
\bibinfo{author}{\bibfnamefont{D.}~\bibnamefont{Braun}} \bibnamefont{and}
  \bibinfo{author}{\bibfnamefont{B.}~\bibnamefont{Georgeot}},
  \bibinfo{journal}{Phys. Rev. A} \textbf{\bibinfo{volume}{72}},
  \bibinfo{pages}{022314} (\bibinfo{year}{2006}).

\bibitem[{\citenamefont{Shi}()}]{Shi02}
\bibinfo{author}{\bibfnamefont{Y.}~\bibnamefont{Shi}},
  \eprint{quant-ph/0205115}.

\bibitem[{\citenamefont{Aharonov}()}]{Aharonov03}
\bibinfo{author}{\bibfnamefont{D.}~\bibnamefont{Aharonov}},
  \eprint{quant-ph/0301040}.

\bibitem[{\citenamefont{Arnaud and Braun}()}]{Arnaud06}
\bibinfo{author}{\bibfnamefont{L.}~\bibnamefont{Arnaud}} \bibnamefont{and}
  \bibinfo{author}{\bibfnamefont{D.}~\bibnamefont{Braun}}, \eprint{to be
  published}.

\bibitem[{\citenamefont{Mehta}(1991)}]{Mehta91}
\bibinfo{author}{\bibfnamefont{M.~L.} \bibnamefont{Mehta}},
  \emph{\bibinfo{title}{{Random Matrices}}} (\bibinfo{publisher}{Academic
  Press}, \bibinfo{address}{New York}, \bibinfo{year}{1991}),
  \bibinfo{edition}{2nd} ed.

\bibitem[{\citenamefont{Prosen et~al.}()\citenamefont{Prosen, Seligman, and
  Weidenmueller}}]{Prosen02}
\bibinfo{author}{\bibfnamefont{T.}~\bibnamefont{Prosen}},
  \bibinfo{author}{\bibfnamefont{T.~H.} \bibnamefont{Seligman}},
  \bibnamefont{and}
  \bibinfo{author}{\bibfnamefont{H.}~\bibnamefont{Weidenmueller}},
  \eprint{math-ph/0203042}.

\bibitem[{\citenamefont{Gorin}(2002)}]{Gorin02}
\bibinfo{author}{\bibfnamefont{T.}~\bibnamefont{Gorin}},
  \bibinfo{journal}{J.Math.Phys} \textbf{\bibinfo{volume}{43}},
  \bibinfo{pages}{3342} (\bibinfo{year}{2002}).

\bibitem[{\citenamefont{Collins and \'Sniady}()}]{Collins04}
\bibinfo{author}{\bibfnamefont{B.}~\bibnamefont{Collins}} \bibnamefont{and}
  \bibinfo{author}{\bibfnamefont{T.}~\bibnamefont{\'Sniady}},
  \eprint{math-ph/0402073}.

\bibitem[{\citenamefont{Aubert and Lam}(2003)}]{Aubert03}
\bibinfo{author}{\bibfnamefont{S.}~\bibnamefont{Aubert}} \bibnamefont{and}
  \bibinfo{author}{\bibfnamefont{C.}~\bibnamefont{Lam}},
  \bibinfo{journal}{J.Math.Phys.} \textbf{\bibinfo{volume}{44}},
  \bibinfo{pages}{6112} (\bibinfo{year}{2003}).

\end{thebibliography}

\end{document}